\documentclass[conference]{IEEEtran}
\usepackage{amsfonts}
\IEEEoverridecommandlockouts

\ifCLASSINFOpdf
\else
\fi
\usepackage{epsfig}
\usepackage{graphicx}
\usepackage{psfig}
\usepackage{epsf}
\usepackage[cmex10]{amsmath}
\usepackage{booktabs}
\usepackage{fancyhdr}
\usepackage{slashbox}
\usepackage{caption2}   
\newtheorem{myTheo}{Theorem}   

\newtheorem{rem}{Remark} 
\begin{document}
\title{UAV-Enabled Mobile Edge Computing: Offloading Optimization and Trajectory Design}
\author{Fuhui Zhou$^{ \S \ddag}$, Yongpeng Wu$^*$, Haijian Sun$^{\S} $, Zheng Chu$^\dagger$ \\
 $^{\S}$Utah State University, USA, $^{\ddag}$Nanchang University, Nanchang, China,\\
$^*$Shanghai Jiao Tong University, China, $^\dagger$University of Surrey, Guildford, U.K\\

Email: \emph{\{zhoufuhui@ieee.org, yongpeng.wu2016@gmail.com, h.j.sun@ieee.org, andrew.chuzheng7@gmail.com\}}
\thanks{The research was supported by the U.S. National Science Foundation grant EARS-1547312, the National Natural Science Foundation of China (61701214, 61701301, 61661028, 61631015, and 61561034), the Young Natural Science Foundation of Jiangxi Province (20171BAB212002), the China Postdoctoral Science Foundation (2017M610400) and the Postdoctoral schedule fund of Jiangxi Province(2017RC17).}}
\maketitle
\begin{abstract}
With the emergence of diverse mobile applications (such as augmented reality), the quality of experience of mobile users is greatly limited by their computation capacity and finite battery lifetime. Mobile edge computing (MEC) and wireless power transfer are promising to address this issue. However, these two techniques are susceptible to propagation delay and loss. Motivated by the chance of short-distance line-of-sight achieved by leveraging unmanned aerial vehicle (UAV) communications, an UAV-enabled wireless powered MEC system is studied. A power minimization problem is formulated subject to the constraints on the number of the computation bits and energy harvesting causality. The problem is non-convex and challenging  to tackle. An alternative optimization algorithm is proposed based on sequential convex optimization. Simulation results show that our proposed design is superior to other benchmark schemes and the proposed algorithm is efficient in terms of the convergence.
\end{abstract}
\IEEEpeerreviewmaketitle
\section{Introduction}
\IEEEPARstart{W}{ITH} the development of Internet of Things (IoT), the emerging diverse mobile applications (augmented reality, face recognition, mobile online gaming, etc.) enable mobile users to enjoy a high quality of experience \cite{R. Q. Hu1}. However, these applications are latency-sensitive and need a high computation capability. Due to the limited battery and low computation capability, it is challenging for mobile devices to execute these applications \cite{F. Zhou4}, \cite{R. Q. Hu}. Fortunately, mobile edge computing (MEC) has been recognized as a promising technique to tackle this challenge \cite{Y. Mao}. It provides the edge network with cloud computing service. Mobile users can offload their computation tasks into the edge network. Unlike mobile cloud computing (MCC), network edge devices in MEC, such as access points, can perform cloud-like computing and are deployed in close proximity to users. MEC has received increasing attention in both academia and industry since it has advantages of saving energy for users, providing low latency services and achieving security for mobile applications \cite{F. Wang}-\cite{C. Wang}.

On the other hand, wireless power transfer (WPT) techniques are promising for prolonging the operational time of energy-limited mobile devices \cite{F. Zhou2}, \cite{X. Lu}. Particularly, radio frequency (RF) signals are used as energy sources for energy harvesting (EH). Compared to the conventional EH techniques, such as solar charging, WPT techniques can provide a controllable and stable power. They are important for energy-limited mobile devices, which are required to execute local low-computation tasks and offload computation-extensive tasks \cite{C. You}-\cite{C. Wang}. However, the harvested power level is greatly influenced by the severe propagation loss. Recently, an unmanned aerial vehicle (UAV)-enabled WPT architecture has been proposed to improve the energy transfer efficiency \cite{J. Xu}. It utilizes an UAV as an energy transmitter for powering the ground mobile users. It was shown that the harvested power level can be significantly improved due to the higher chance of short-distance line-of-sight (LoS) energy transmit links \cite{J. Xu2}.

Motivated by the UAV-enabled WPT architecture, a UAV-enabled wireless powered MEC system is studied in this paper. In the system, the UAV transmits energy to multiple ground users and the ground users exploit the harvested energy for local computing and computation tasks offloading. \emph{To the authors' best knowledge, this is the first work that establishes an UAV-enabled wireless powered MEC system and studies the joint optimization of computation offloading and trajectory design}. The related works are summarized as follows.

In \cite{F. Wang}, the revenue of the wireless cellular networks with MEC was maximized by jointly optimizing the computation offloading decision and resource allocation. The authors in \cite{C. You}-\cite{C. Wang} extended the resource allocation problems into wireless powered MEC systems. Specifically, in \cite{C. You}, an energy-efficient resource allocation strategy was proposed by jointly optimizing the number of the local computation bits and the offloading computation bits under the causal energy harvesting constraint. The authors in \cite{J. Xu33} proposed an efficient reinforcement learning-based resource management scheme in an MEC system with energy harvesting. It was shown that the computation performance can be significantly improved by using the proposed algorithm. In \cite{S. Bi} and \cite{C. Wang}, a joint optimization framework was proposed in wireless powered MEC systems with different operational paradigms, namely, binary and partial offloading, respectively. In the binary offloading paradigm, the computation task is completely executed in mobile devices or offloaded into network edge devices for computing. In the partial offloading paradigm, the computation task can be divided into two parts, one for local computing and the other for offloading. The computation rate was maximized in \cite{S. Bi} and the transmit power was minimized in \cite{C. Wang} by jointly optimizing the computing frequency and the transmit power.

In \cite{C. You}-\cite{C. Wang}, the access point or the energy transmitter equipped with an MEC server is deployed at the fixed location. It results in a low energy transfer efficiency due to the severe propagation loss \cite{F. Zhou2}, \cite{X. Lu}. In order to tackle this issue, the authors designed UAV-enabled wireless powered systems and jointly optimized the resource allocation and trajectory of the UAV \cite{J. Xu}, \cite{J. Xu2}. However, MEC was not considered in \cite{J. Xu} and \cite{J. Xu2}. Recently, an UAV-enabled MEC system was designed in \cite{S. Jeong} and the transmit power of users was minimized by jointly optimizing the number of the local computation bits, the offloading computation bits and the downloading bits.

Different from \cite{J. Xu}-\cite{S. Jeong}, an UAV-enabled wireless powered MEC system is studied in this paper. A power minimization problem is formulated by jointly optimizing the number of the offloading computation bits, the local computation frequencies of users and the UAV, and the trajectory of the UAV. It is challenging to solve the formulated non-convex problem due to the existing couple among the optimized variables. An alternative optimization algorithm is proposed to solve it by using the sequential convex optimization (SCA) techniques. Simulation results show that our proposed resource allocation scheme outperforms other benchmark schemes.

The remainder of this paper is organized as follows. The system model is presented in Section II. Section III presents the energy minimization problem. Section IV presents simulation results. The paper concludes with Section V.
\section{System Model}
\begin{figure}[!t]
\centering
\includegraphics[width=2.8 in]{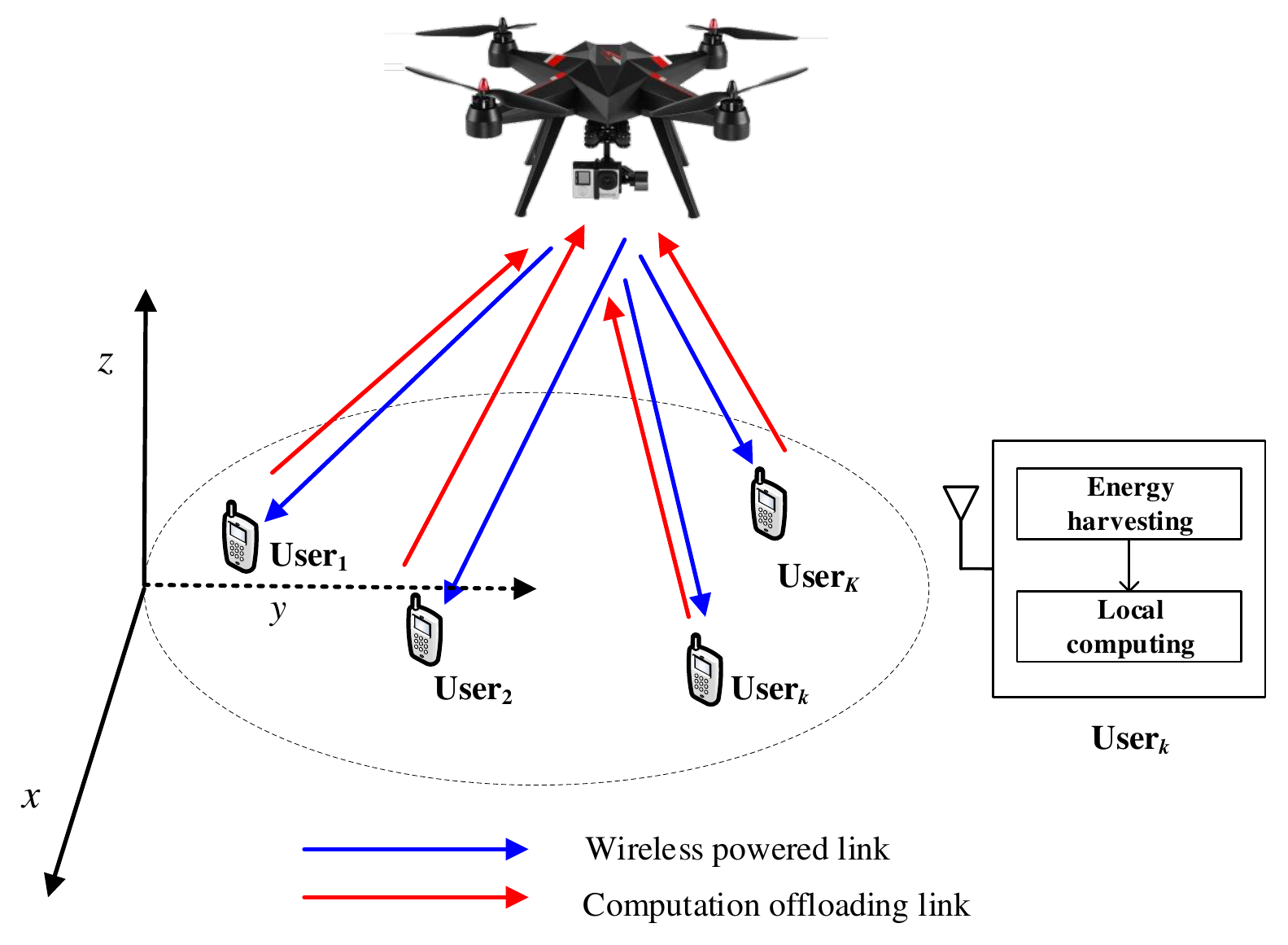}
\caption{The system model.} \label{fig.1}
\end{figure}
An UAV-enabled wireless powered MEC system is considered in Fig. 1, where an UAV equipped with an MEC server transmits energy to $K$ users and provides MEC services for these users. In this paper, the partial offloading paradigm is applied. Similar to \cite{S. Bi} and \cite{C. Wang}, users can simultaneously perform energy harvesting, local computing and computation offloading. Without loss of generality, a three-dimensional
(3D) Euclidean coordinate is adopted. Each user is fixed at the ground. The location of the $k$th ground user is denoted by $\mathbf{q}_k$, where $\mathbf{q}_k=[x_k, y_k]$, $k\in {\cal K}$ and $ {\cal K}=\left\{1, 2, \cdots, K\right\}$. Boldface lower case letters represent vectors, and $x_k$ and $y_k$ are the horizontal plane coordinate of the $k$th ground user. It is assumed that the positions of users are known to the UAV for designing trajectory. A  finite time horizon with during $T$ is considered. During the finite time, the UAV flies at a fixed altitude ($H>0$). A block fading channel model is applied. During the finite time, the channel is unchanged.

For ease of exposition, the finite time $T$ is discretized into $N$ equal-time slots, denoted by $n= 1, 2, \cdots, N$. At the $n$th slot, it is assumed that the horizontal plane coordinate of the UAV is $\mathbf{q}_u\left[n\right]=[x_u[n], y_u[n]]$. Similar to \cite{J. Xu2}-\cite{Y. Zeng}, it is assumed that the wireless channel between the UAV and each user is dominated by the LoS channel. Thus, the channel power gain between the UAV and the $k$th user is denoted by ${h_k}\left[ n \right]$, given as
\begin{align}\label{27}\
{h_k}\left[ n \right] = {\beta _0}d_{k,n}^{ - 2} = \frac{{{\beta _0}}}{{{H^2} + {{\left\| {{\mathbf{q}_u}\left[ n \right] - {\mathbf{q}_k}} \right\|}^2}}}, k\in {\cal K}, n\in {\cal N},
\end{align}
where $\beta _0$ is the channel power gain at a reference distance $d_0=1$ m; $d_{k,n}$ is the horizontal plane distance between the UAV and the $k$th user at the $n$th slot, $n\in {\cal N}$, ${\cal N}=\left\{1, 2, \cdots, N\right\}$, and $\left\| \cdot\right\|$ denotes its Euclidean norm. In order to reach meaningful insights into the design of an UAV-enabled wireless powered MEC system, similar to \cite{C. You}-\cite{F. Zhou2}, a linear EH model is applied. Thus, the harvested energy at the $k$th user during $n$ time slots denoted by ${E_k}\left[ n \right]$, is given as
\begin{align}\label{27}\
{E_k}\left[ n \right] = \sum\limits_{i = 1}^n {\frac{{T\eta {h_k}\left[ i \right]{P_u}}}{N}},
\end{align}
where $\eta$ denotes the energy conservation efficiency, $0<\eta\leq1$ and ${P_u}$ is the transmit power of the UAV. In this paper, the UAV employs a constant power transmission \cite{J. Xu}-\cite{Y. Zeng}. During the $n$th slot, all users perform energy harvesting, local computing and computation offloading. In order for all users to offload their bits to the UAV for computation, a time division multiple access (TDMA) protocol is applied. The time interval $T/N$ is divided into $K$ time slots with duration $\lambda= T/(NK)$ and $K$ users offload their computation bits to the UAV one by one. Similar to \cite{C. You}-\cite{C. Wang}, the received energy and the energy for transmitting the computed results of the UAV are ignored.

Let ${{l_k}\left[ n \right]}$ and  ${{f_k}\left[ n \right]}$ denote the number of the offloading bits and the central processing unit (CPU) frequency (cycles/s) of the $k$th user at the $n$th slot, respectively. Thus, the transmit power of the $k$th user for offloading ${{l_k}\left[ n \right]}$ computation bits denoted by $P_k[n]$, is given as
\begin{align}\label{27}\
{P_k}\left[ n \right] = \frac{{\Gamma {\sigma ^2}\left( {{2^{\frac{{{l_k}\left[ n \right]}}{{B\lambda }}}} - 1} \right)}}{{{h_k}\left[ n \right]}},
\end{align}
where $B$ is the communication bandwidth and $\sigma ^2$ denotes the noise power at the user. In $\left(3\right)$, $\Gamma$ is a constant related to the gap from the channel capacity owning to a practical coding and modulation scheme. It is assumed that $\Gamma=1$ in this paper for simplicity. Let ${{f_u}\left[ n \right]}$ denote the CPU frequency of the UAV at the $n$th slot. According to \cite{C. You}-\cite{C. Wang}, the energy consumed for the local computation at the $k$th user and that for the offloading computation at the UAV in the $n$th slot are denoted by ${E_{k,l}}\left[ n \right]$ and ${E_{u,o}}\left[ n \right]$, respectively given as
\begin{subequations}
\begin{align}\label{27}\
&{E_{k,l}}\left[ n \right] = {\gamma _c}K\lambda {\left[ {{f_k}\left[ n \right]} \right]^3},\\
&{E_{u,o}}\left[ n \right] = {\gamma _c}K\lambda {\left[ {{f_u}\left[ n \right]} \right]^3},
\end{align}
\end{subequations}
where ${\gamma _c}$ is the effective switched capacitance of the CPU. Similar to the works in \cite{S. Jeong} and \cite{N. Xue}, the propulsion energy  consumption model at the UAV due to the flying in the $n$th slot denoted by $E_s\left[ n \right]$, is given as
\begin{subequations}
\begin{align}\label{27}\
&{E_{s}}\left[ n \right] =  {\kappa {{\left\| {{v_u}\left[ n \right]} \right\|}^2} },\\
&{{v_u}\left[ n \right]}=\frac{{\left\| {{\mathbf{q}_u}\left[ {n + 1} \right] - {\mathbf{q}_u}\left[ n \right]} \right\|}}{{K\lambda }},
\end{align}
\end{subequations}
where $\kappa  = 0.5WT/N$ and $W$ is the mass of the UAV. Note that the propulsion energy consumption model employed in this paper only depends on the velocity. In future work, we will exploit a more general model that considers both the velocity and acceleration of the UAV.
\section{Energy Minimization Design}
\subsection{The Energy Minimization Problem Formulation}
In the UAV-enabled wireless powered MEC system, in order to minimize the energy consumed at the UAV while guaranteeing the computation bits of all users, the number of the offloading computation bits and the CPU frequency of the users, the CPU frequency and the trajectory of the UAV are jointly optimized. The energy minimization problem can be formulated as $\text{P}_{{1}}$, given as
\begin{subequations}
\begin{align}\label{27}\
&\text{P}_{{1}}: {\mathop {\min }\limits_{\Xi} }\ {\sum\limits_{n = 1}^N { {\kappa {{\left\| {{v_u}\left[ n \right]} \right\|}^2} } }  + {{T{P_u}}}+ \sum\limits_{n = 2}^N {{\gamma _c}K\lambda {{\left[ {{f_u}\left[ n \right]} \right]}^3}} },\\
&\text{s.t.} \ \  C1:\sum\limits_{n = 1}^N {\frac{{\lambda K{f_k}\left[ n \right]}}{M}}  + \sum\limits_{n = 1}^{N - 1} {{l_k}\left[ n \right]}  = {R_k},k\in {\cal K},\\
& C2:\sum\limits_{i = 1}^n {{E_{k,l}}\left[ i \right]}  + \sum\limits_{i = 1}^n {\lambda {P_k}\left[ i \right]}  \le \sum\limits_{i = 1}^n {{E_k}\left[ i \right]} ,k\in {\cal K},n\in {\cal N},\\
&  C3:\sum\limits_{j = 2}^n {\frac{{{f_u}\left[ j \right]K\lambda }}{M}}  \le \sum\limits_{k = 1}^K {\sum\limits_{i = 1}^{n - 1} {{l_k}\left[ i \right]} }, k\in {\cal K},n\in {\cal N}_{-N},\\
&  C4: \sum\limits_{j = 2}^N {\frac{{{f_u}\left[ j \right]K\lambda }}{M}}  = \sum\limits_{k = 1}^K {\sum\limits_{i = 1}^{N - 1} {{l_k}\left[ i \right]} }, k\in {\cal K},\\
&  C5: {l_k}\left[ N \right] = 0,{f_u}\left[ 1 \right] = 0, k\in {\cal K},\\
& C6: {\left\| {\mathbf{q}_u\left[ {n + 1} \right] - \mathbf{q}_u\left[ n \right]} \right\|} \le {V_{\max }}K\lambda, n\in {\cal N},\\
& C7: \mathbf{q}_u\left[ 1 \right] = {\mathbf{q}_0},\mathbf{q}_u\left[ {N + 1} \right] = {\mathbf{q}_F},\\
& C8: f_u\left[ n \right]\geq0,f_k\left[ {n} \right] \geq0, k\in {\cal K},n\in {\cal N},
\end{align}
\end{subequations}
where $\Xi$ denotes the variable set consisting of ${f_u}\left[ n \right],\mathbf{q}_u\left[ n \right],
{l_k}\left[ n \right],{f_k}\left[ n \right]
$; ${R_k}$ denotes the total number of the computation bits of the $k$th user; $V_{\max }$ is the maximum flying speed of the UAV; ${\mathbf{q}_0}$ and ${\mathbf{q}_F}$ are the destined initial and final locations of the UAV, respectively; $M$ denotes the number of CPU cycles required for computing one bit at the user and the UAV. ${\cal N}_{-N}$ denotes the set ${\cal N}$ other than $N$. The constraint $C1$ is the total computation bits required at the $k$th user; $C2$ is the energy causal constraint that the energy consumed for the local computation and offloading computation bits cannot be higher than the harvesting energy; $C3$ represents that the number of the computation bits at the UAV in the $n$th slots cannot be higher than the total number of the offloading computation bits of all users before the $n-1$th slot. Note that the UAV starts to compute the offloading bits at the $n$th slots only when all users finish offloading the computation bits of the $n-1$th slot; $C4$ denotes that all the offloading computation bits of users should be computed; $C5$ represents that the UAV does not execute the computation task in the first slot and all users do not offload their computation tasks in the last slot; $C6$ is the flying speed constraint and $C7$ is the initial and final locations constraint related to the UAV.

It is challenging to solve the non-convex problem $\text{P}_{{1}}$ due to the presence of the couple among the optimization variables $\mathbf{q}_u\left[ n \right]$, ${l_k}\left[ n \right]$ and ${f_k}\left[ n \right] $. An alternative algorithm is proposed to solve $\text{P}_{{1}}$ in the following subsection.

\subsection{Computation Offloading And CPU Frequency Optimization}
It is seen from $\text{P}_{{1}}$ that $\text{P}_{{1}}$ is convex for a given trajectory $\mathbf{q}_u\left[ n \right]$. Thus, for a given $\mathbf{q}_u\left[ n \right]$, $\text{P}_{{1}}$ can be transformed as $\text{P}_{{2}}$, given as
\begin{subequations}
\begin{align}\label{27}\
&\text{P}_{{2}}:\ {\mathop {\min }\limits_{{\scriptstyle{f_u}\left[ n \right],
\scriptstyle{l_k}\left[ n \right],{f_k}\left[ n \right]\hfill}} }\ { \sum\limits_{n = 2}^N {{\gamma _c}K\lambda {{\left[ {{f_u}\left[ n \right]} \right]}^3}} },\\
&\text{s.t.} \ \  C1-C5 \ \text{and}\ C8.
\end{align}
\end{subequations}
Since $\text{P}_{{2}}$ is convex, it can be solved by using the Lagrange duality method \cite{S. P. Boyd}, \cite{Q. Li}. By solving $\text{P}_{{2}}$, Theorem 1 can be obtained as follows.
\begin{myTheo}
For a given trajectory $\mathbf{q}_u\left[ n \right]$, the optimal offloading computation bits and the CPU frequency of the users, and the CPU frequency of the UAV denoted by ${l_k}^{opt}\left[ n \right], {f_u}^{opt}\left[ n \right]$ and ${f_k}^{opt}\left[ n \right]$,  can be respectively given as
\begin{subequations}
\begin{align}\label{27}\
&{l_k}^{opt}\left[ n \right]{\rm{ = }}{B\lambda {\rm{lo}}{{\rm{g}}_2}\left\{ {\frac{{B{h_k}\left[ n \right]\left[ {\sum\limits_{j = n{\rm{ + }}1}^{N{\rm{ - }}1} {{\theta _j}}  + {\mu _k} - {\theta _N}} \right]}}{{\sum\limits_{j = n}^N {{\nu _{k,j}}} \Gamma {\sigma ^2}\ln 2}}} \right\}},\\
&{f_u}^{opt}\left[ n \right] = \left\{ \begin{array}{l}
\begin{array}{*{20}{c}}
{0,}&{n = 1}
\end{array}\\
\begin{array}{*{20}{c}}
{\sqrt {\frac{{{\theta _N} - \sum\limits_{j = n}^{N - 1} {{\theta _j}} }}{{{3\gamma _c}M}},} }&{n = 2, \cdots ,N - 1}
\end{array}\\
\begin{array}{*{20}{c}}
{\sqrt {\frac{{{\theta _N}}}{{{3\gamma _c}M}},} }&{n = N}
\end{array}
\end{array} \right.
\\
&{f_k}^{opt}\left[ n \right] = \sqrt {\frac{{{\mu _k}}}{{3{\gamma _c}MK\sum\limits_{j = n}^N {{\nu _{k,j}}} }}}, k\in {\cal K},n\in {\cal N}
\end{align}
\end{subequations}
where $\mu _k\geq0$, $\nu _{k,n}\geq0$ and $\theta _n\geq0$ are the dual variables associated with the constraints $C1$, $C2$, $C3$ and $C4$, respectively.
\end{myTheo}
\begin{IEEEproof}
See Appendix A.
\end{IEEEproof}
\begin{rem}
Theorem 1 indicates that the CPU frequency of the UAV increases with the time slots since $\theta _N>0$ and $\theta _n\geq0$ when $n=2,3,\cdots, N-1$. It means that the total number of the offloading computation bits increases with the time slots. Thus, in order to decrease the total energy consumed at the UAV, users need to allocate a high energy for local computation so that the number of the  offloading computation can be decreased. It is also seen that the number of the offloading computation bits is increased when the channel condition between the UAV and users is improved. This indicates that the number of the offloading computation bits of users increases with the decrease of the distance between the user and the UAV. Finally, the dual variables can be obtained by using the subgradient algorithm \cite{F. H. Zhou1}.
\end{rem}

\subsection{Trajectory Optimization}
For any given number of the offloading computation bits, the CPU frequencies of users and the UAV, the trajectory optimization problem can be formulated as $\text{P}_{{3}}$, given as
\begin{subequations}
\begin{align}\label{27}\
&\text{P}_{{3}}:\ {\mathop {\min }\limits_{\mathbf{q}_u\left[ n \right]} }\ {\sum\limits_{n = 1}^N {\kappa {{\left\| {{v_u}\left[ n \right]} \right\|}^2}} }\\
&\text{s.t.} \ \  C2, C6 \ \text{and}\ C7.
\end{align}
\end{subequations}
Due to the constraint $C2$,  $\text{P}_{{3}}$ is non-convex. In order to tackle $C2$, the SCA technique is exploited. It can guarantee that the obtained solutions satisfy the Karush-Kuhn-Tucker (KKT) conditions of $\text{P}_{{3}}$. By using the SCA technique, Theorem 2 is given as follows.
\begin{myTheo}
For any local trajectory $\mathbf{q}_u^{j}\left[ n \right], n\in {\cal N}$ at the $j$th iteration, one has
\begin{subequations}
\begin{align}\label{27}\
&\sum\limits_{i = 1}^n {\frac{{K\lambda \eta {P_u}{\beta _0}}}{{{H^2} + {{\left\| {{\mathbf{q}_u}\left[ i \right] - {\mathbf{q}_k}} \right\|}^2}}}}  \ge K\lambda \eta {P_u}{\beta _0}\overline {{h_k}} \left[ n \right],\\
&\overline {{h_k}} \left[ n \right] = \sum\limits_{i = 1}^n\left\{ {\frac{{{H^2} + 2{{\left\| {\mathbf{q}_u^j\left[ i \right] - {\mathbf{q}_k}} \right\|}^2} - {{\left\| {{\mathbf{q}_u}\left[ i \right] - {\mathbf{q}_k}} \right\|}^2}}}{{{{\left( {{H^2} + {{\left\| {\mathbf{q}_u^j\left[ i \right] - {\mathbf{q}_k}} \right\|}^2}} \right)}^2}}}}\right\},
\end{align}
\end{subequations}
where the equality holds when $\mathbf{q}_u\left[ n \right]=\mathbf{q}_u^j\left[ n \right] $.
\end{myTheo}
\begin{IEEEproof}
Let $f\left( z \right) = \frac{a}{{b + z}}$, where $a$ and $b$ are positive constants, and $z\geq0$. Since $f\left( z \right)$ is convex with respect to $z$, the following inequality can be obtain:
\begin{align}\label{27}\
\frac{a}{{b + z}} \ge \frac{a}{{b + {z_0}}} - \frac{a}{{{{\left( {b + {z_0}} \right)}^2}}}\left( {z - {z_0}} \right),
\end{align}
where $z_0$ is a given local point. By using eq. $\left(11\right)$, Theorem 2 is obtained.
\end{IEEEproof}

Using Theorem 2, $\text{P}_{{3}}$ can be solved by iteratively solving the approximate problem $\text{P}_{{4}}$, given as
\begin{subequations}
\begin{align}\label{27}\
&\text{P}_{{4}}:\ {\mathop {\min }\limits_{\mathbf{q}_u\left[ n \right]} }\ {\sum\limits_{n = 1}^N {\kappa {{\left\| {{v_u}\left[ n \right]} \right\|}^2}} },\\
&\text{s.t.} \ \ \  C6 \ \text{and}\ C7,\\
& \sum\limits_{i = 1}^n {{E_{k,l}}\left[ i \right]}  + \sum\limits_{i = 1}^n {\lambda {P_k}\left[ i \right]}  \le  K\lambda \eta {P_u}{\beta _0}\overline {{h_k}} \left[ n \right] ,k\in {\cal K},n\in {\cal N}.
\end{align}
\end{subequations}
It is seen that $\text{P}_{{4}}$ is convex and can be readily solved by using the software \texttt{CVX} \cite{F. Zhou2}. Based on solving $\text{P}_{{2}}$ and $\text{P}_{{4}}$, an alternative optimization algorithm denoted by Algorithm 1 is given to solve $\text{P}_{{1}}$. The details for Algorithm 1 can be found in Table 1. In Table 1, $E_u^{i}$ denotes the value of the objective function of $\text{P}_{{1}}$.
\begin{figure*}[!t]
\normalsize
\begin{align}\label{27}\ \notag
L\left( {{\Xi _1}} \right) =& \sum\limits_{n = 2}^N {{\gamma _c}K\lambda {{\left[ {{f_u}\left[ n \right]} \right]}^3}}  + \sum\limits_{k = 1}^K {{\mu _k}} \left[ {{R_k} - \sum\limits_{n = 1}^N {\frac{{\lambda {f_k}\left[ n \right]}}{M}}  - \sum\limits_{n = 1}^{N - 1} {{l_k}\left[ n \right]} } \right]\\  \notag
 &+ \sum\limits_{k = 1}^K {\sum\limits_{n = 1}^N {{\nu _{k,n}}\left\{ {\sum\limits_{i = 1}^n {{E_{k,l}}} \left[ i \right] + \sum\limits_{i = 1}^n {\lambda {P_k}\left[ i \right]}  - \sum\limits_{i = 1}^n {{E_k}\left[ i \right]} } \right\}} }  + \sum\limits_{n = 1}^{N - 1} {{\theta _n}\left\{ {\sum\limits_{j = 2}^n {\frac{{{f_u}\left[ j \right]K\lambda }}{M}}  - \sum\limits_{k = 1}^K {\sum\limits_{i = 1}^{n - 1} {{l_k}\left[ i \right]} } } \right\}} \\
& + {\theta _N}\left\{ {\sum\limits_{k = 1}^K {\sum\limits_{i = 1}^{N - 1} {{l_k}\left[ i \right]}  - \sum\limits_{j = 2}^N {\frac{{{f_u}\left[ j \right]K\lambda }}{M}} } } \right\} + \sum\limits_{k = 1}^K {{\rho _k}{l_k}\left[ N \right]}  + \vartheta {f_u}\left[ 1 \right]
\end{align}
\hrulefill \vspace*{4pt}
\end{figure*}
\begin{table}[htbp]
\begin{center}
\caption{The alternative optimization algorithm}
\begin{tabular}{lcl}
\\\toprule
$\textbf{Algorithm 1}$: The alternative optimization algorithm for $\text{P}_{{1}}$\\ \midrule
\  1: \textbf{Setting:}\\
\ \  \ $R_k$, $k\in  {\cal K}$, $P_u$, $T$, $N$, ${V_{\max }}$, $\mathbf{q}_0$, $\mathbf{q}_F$, and the tolerance errors $\xi$,  $\xi_1$; \\
\  2: \textbf{Initialization:}\\
\ \  \ The iterative number $i=1$ and $\mathbf{q}_u^i\left[ n \right]$; \\
\  3: \textbf{Repeat 1:}\\
 \   \ \ \ \ \ calculate ${l_k}^{opt,i}\left[ n \right], {f_u}^{opt,i}\left[ n \right]$ and ${f_k}^{opt,i}\left[ n \right]$ \\
 \   \ \ \ \ \  using eq. $\left(8\right)$ for given  $\mathbf{q}_u^i\left[ n \right]$; \\
\ \ \ \ \ \ update $\mu _k$, $\nu _{k,n}$ and $\theta _n$ using the subgradient algorithm; \\
 \ \ \ \ \ \ initialize the iterative number $j=1$; \\
 \ \ \ \ \ \ \textbf{Repeat 2:}\\
  \ \ \ \ \ \ \ \ solve $\text{P}_{\textbf{4}}$ by using \texttt{CVX} for the given ${l_k}^{opt,i}\left[ n \right]$, \\
  \ \ \ \ \ \ \ \ ${f_u}^{opt,i}\left[ n \right]$ and ${f_k}^{opt,i}\left[ n \right]$;  \\
  \ \ \ \ \ \ \ \ update $j=j+1$, and $\mathbf{q}_u^j\left[ n \right]$;\\
 \ \ \ \ \ \ \ \  if $\sum\limits_{n = 1}^N {\left\| {\mathbf{q}_u^j\left[ n \right] - \mathbf{q}_u^j\left[ n \right]} \right\|}  \le \xi$   \\
  \ \ \ \ \ \ \ \  \ \ \ $\mathbf{q}_u^i\left[ n \right]=\mathbf{q}_u^j\left[ n \right]$ ;\\
   \ \ \ \ \ \ \ \  \ \ \ break; \\
 \ \ \ \ \ \ \ \  end \\
  \ \ \ \ \ \  \textbf{end Repeat 2} \\
 \ \ \ \ \ update the iterative number $i=i+1$;  \\
\ \ \ \ \ if $\left|E_u^{i}-E_u^{i-1}\right|\leq \xi_1$ \\
\ \ \ \ \  \  break;\\
\ \ \ \ \  end\\
\ \ \ \ \textbf{end Repeat 1}\\
\  4: \textbf{Obtain solutions:}\\
 \ \ \ \ \ \ ${l_k}^{opt}\left[ n \right]$, ${f_u}^{opt}\left[ n \right]$ and ${f_k}^{opt}\left[ n \right]$ and $\mathbf{q}_u^{opt}\left[ n \right]$. \\
\bottomrule
\end{tabular}
\end{center}
\end{table}

\section{Simulation Results}
In this section, simulation results are presented to compare the performance obtained by using our proposed design with that achieved by using two benchmark schemes, denoted by Scheme 1 and Scheme 2, respectively.  In Scheme 1,  the UAV flies straight with a constant speed from the initial position to the final position. In Scheme 2, the UAV flies along the trajectory that is a semi-circle with its diameter being ${\left\| {\mathbf{q}_F- {\mathbf{q}_0}} \right\|}$. The converge performance of the proposed algorithm is also evaluated by simulation results. The simulation settings are based on the works in  \cite{C. Wang} and \cite{S. Jeong}. The positions and the total number of the computation bits of users are set as: $\mathbf{q}_1=[0,0]$, $\mathbf{q}_2=[0,10]$, $\mathbf{q}_3=[10,10]$, $\mathbf{q}_4=[10,0]$, $R_1=2$ Mbits, $R_2=4$ Mbits, $R_3=6$ Mbits, and $R_4=3$ Mbits, respectively. The detail settings are given in Table II.
\begin{table}[htbp]
 \caption{\label{tab:test}Simulation Parameters}
 \begin{tabular}{l|c|c}
  \midrule
  \midrule
  Parameters & Notation & Typical Values  \\
  \midrule
  \midrule
 Numbers of Users & $K$ & $4$ \\
The height of the UAV & $H$ & $10$ m \\
The time length of the UAV flying & $T$ & $2$ sec \\
Numbers of  CPU cycles & $M$ & $10^3$ cycles/bit \\
Energy conversation efficiency & $\eta$ & $0.8$ \\
Communication bandwidth & $B$ & $40$ MHz \\
The receiver noise power & $\sigma^2$ & $10^{-9}$ W\\
The number of time slots & $N$ & $50$  \\
The mass of the UAV& $W$ & $9.65$ kg \\
The effective switched capacitance & $\gamma_c$ & $10^{-28}$ \\
The channel power gain &$\beta_0$& $-50$ dB\\
 The tolerance error & $\xi, \xi_1$ & $10^{-4}$ \\
 The initial position of the UAV & $\mathbf{q}_0$ & $[0, 0]$ \\
  The final position of the UAV & $\mathbf{q}_F$ & $[10, 0]$ \\
The maximum speed of the UAV & $V_{\max }$  & $10$ m/s\\
The transmit power of the UAV & $P_u$  & $100$ dBm\\
\midrule
 \end{tabular}
\end{table}

\begin{figure}[!t]
\centering
\includegraphics[width=2.4 in]{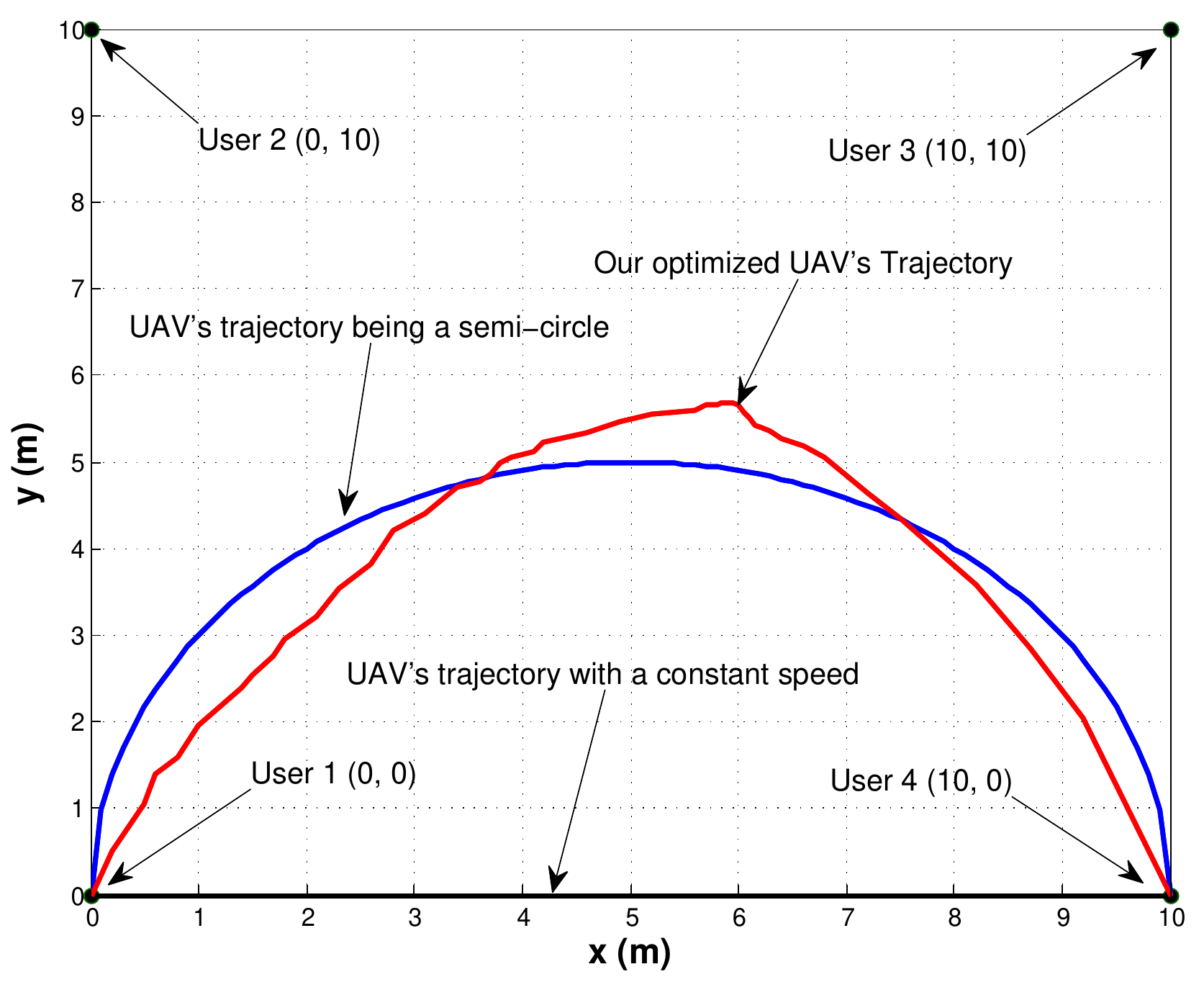}
\caption{The trajectories of the UAV under different schemes with $T=2$ seconds.} \label{fig.1}
\end{figure}
Fig. 2 shows the trajectories of the UAV under different schemes. The time length of the UAV flying is set as $T=2$ seconds. The trajectories of the UAV under Scheme 1 and Scheme 2 are also presented. As shown in Fig. 2, under our proposed optimal trajectory, the UAV firstly flies smoothly and tends to User 2 and User 3, and then the UAV flies smoothly with a higher speed to the final position. The reason is that the UAV needs to provide more energy to User 2 and  User 3, which has a larger number of computation bits to be offloaded. Moreover, in order to control the total number of the computation bits of all users offloaded to the UAV, the UAV flies with a higher speed in the end of the flying time so that the harvested energy of users used for offloading the computation bits can be compromised.

\begin{figure}[!t]
\centering
\includegraphics[width=2.4 in]{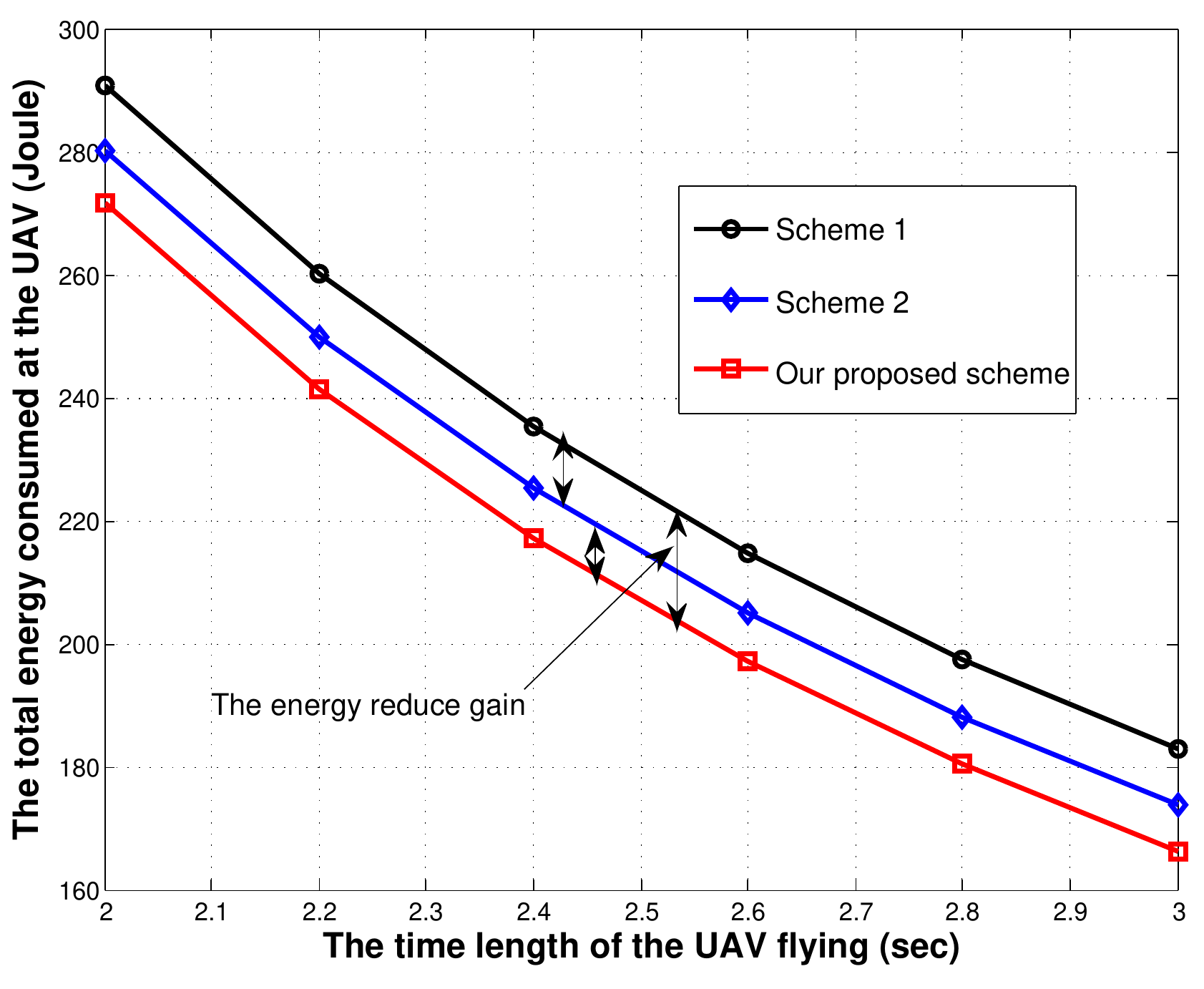}
\caption{The total energy consumed at the UAV versus the time length of the UAV flying under different schemes.} \label{fig.1}
\end{figure}
Fig. 3 shows the total energy consumed at the UAV versus the time length of the UAV flying under different schemes. It is seen that the energy consumed at the UAV by using our proposed scheme is the smallest among those by using the benchmark schemes. This demonstrates that our proposed scheme that jointly optimizes the number of the offloading computation bits, the CPU frequency of users and the UAV, and the trajectory of the UAV can is more efficient in terms of the energy minimization of the UAV. It is also seen that the total energy consumed at the UAV decreases with the increase of the time length of the UAV flying, irrespective of the adopted scheme. It can be explained by the fact that the total energy consumed at the UAV is dominated by the flying speed and the CPU frequency of the UAV, and the flying speed and the CPU frequency can be decreased when the flying time is increased.

\begin{figure}[!t]
\centering
\includegraphics[width=2.4 in]{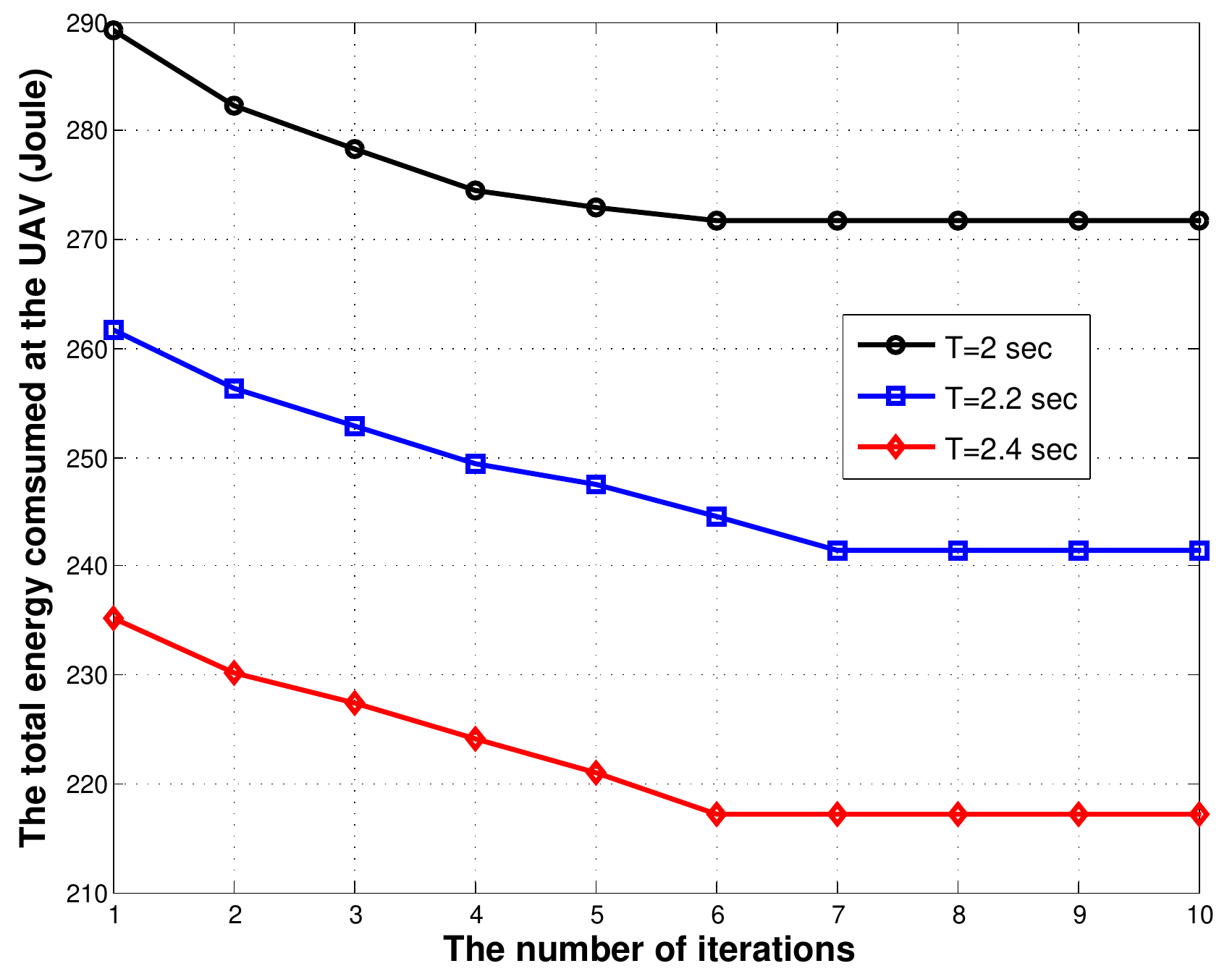}
\caption{The total energy consumed at the UAV versus the number of iterations $T=2, 2.2 $ or $2.4$ seconds.} \label{fig.1}
\end{figure}
Fig. 4 is presented to verify the efficiency of our proposed alternative algorithm. It can be seen that only several number of iterations are required for Algorithm 1 to converge.
\section{Conclusion}
An UAV-enabled wireless powered MEC system was studied where the UAV provides multiple ground users with computation
offloading and sustainable operation opportunities. The number of the offloading computation bits and the CPU frequency of users, the CPU frequency and the trajectory of the UAV were jointly optimized in order to minimize the energy consumed at the UAV. An alternative algorithm was proposed based on the SCA techniques. Simulation results show that our proposed design outperforms other benchmark schemes and that the proposed algorithm only requires several number of iterations to converge.
\appendices
\section{Proof of Theorem 1}
The Lagrangian of $\text{P}_{{2}}$ related to the proof is given by eq. $\left(13\right)$ at the top of the previous page, where ${\rho _k}\geq0$ and $\vartheta\geq0$ are the dual variables related to the constraint $C5$; $\Xi _1$ is the set consisting of all optimization and dual variables. Thus, the derivations of the Lagrangian of $\text{P}_{{2}}$ with respect to  ${l_k}\left[ n \right]$ and ${f_k}\left[ n \right]$, can be respectively given as

\begin{subequations}
\begin{align}\label{27}\ \notag
&\frac{{\partial L\left( \Xi  \right)}}{{\partial {l_k}\left[ n \right]}}\\
&= \frac{{\Gamma {\sigma ^2}\left( {{2^{\frac{{{l_k}\left[ n \right]}}{{B\lambda }}}}} \right)\ln 2}}{{B{h_k}\left[ n \right]}}\sum\limits_{j = n}^N {{\nu _{k,j}}}  - \left[ {\sum\limits_{j = n{\rm{ + }}1}^{N - 1} {{\theta _j}}  + {\mu _k} - {\theta _N}} \right], \\
&\frac{{\partial L\left( \Xi  \right)}}{{\partial {f_k}\left[ n \right]}} =  - \frac{{{\mu _k}\lambda }}{M} + 3{\gamma _c}K\lambda \sum\limits_{j = n}^N {{\nu _{k,j}}{{\left[ {{f_k}\left[ n \right]} \right]}^2}}.
\end{align}
\end{subequations}
Let their derivations be zero. Thus, eq. $\left(8\rm{a}\right)$ and eq. $\left(8\rm{c}\right)$ are obtained. Let the derivation of the Lagrangian of $\text{P}_{{2}}$ with respect to ${f_u}\left[ n \right]$ be zero. One has
\begin{subequations}
\begin{align}\label{27}\
&3{\gamma _c}{\left[ {{f_u}\left[ n \right]} \right]^2}{\rm{ + }}\frac{{\left( {\sum\limits_{j = n}^{N - 1} {{\theta _j}}  - {\theta _N}} \right)}}{M}{\rm{ = }}0,n = 2, \cdots ,N - 1, \\
&3{\gamma _c}{\left[ {{f_u}\left[ N \right]} \right]^2}{\rm{ - }}\frac{{{\theta _N}}}{M}{\rm{ = }}0,n = N.
\end{align}
\end{subequations}
From eq. $\left(15\right)$, eq. $\left(8\rm{b}\right)$ is obtained. The proof for Theorem 1 is completed.

\end{document}